\documentclass[prl,twocolumn,aps,showpacs,10pt]{revtex4-1}

\usepackage{graphicx}  
\graphicspath{{./Figures/}}
\usepackage{dcolumn}  
\usepackage{amssymb, amsmath}
\usepackage{hyperref}

\begin{document}

\title{Dipolar and scalar $^{3}$He-$^{129}$Xe frequency shifts in mm-sized stemless cells}
\author{M.\ E.\ Limes}
\author{N.\ Dural}
\author{M.\ V.\ Romalis}
\affiliation{Department of Physics, Princeton University, Princeton, New Jersey, 08544, USA}
\author{E.\ L.\ Foley}
\author{T.\ W.\ Kornack}
\author{A. Nelson}
\author{L.\ R.\ Grisham}
\affiliation{Twinleaf LLC, Princeton, New Jersey, 08544, USA}
\date{\today}
\begin{abstract}

We describe a $^{3}$He-$^{129}$Xe comagnetometer operating in stemless anodically bonded cells with a 6~mm$^3$ volume and a $^{129}$Xe spin coherence time of 300 s.   We use a  $^{87}$Rb pulse-train magnetometer with co-linear pump and probe beams to study the nuclear spin frequency shifts caused by spin polarization of $^{3}$He. By systematically varying the cell geometry in a batch cell fabrication process we can separately measure the cell shape dependent and independent frequency shifts. We find that a certain aspect ratio  of the cylindrical  cell can  cancel the effects of $^3$He magnetization that limit the stability of vapor-cell comagnetometers. Using this control we also observe for the first time a scalar $^{3}$He-$^{129}$Xe collisional frequency shift characterized by an enhancement factor $\kappa_{\text{HeXe}} = -0.011\pm0.001$.

\end{abstract}
\pacs{32.30.Dx, 06.30.Gv,39.90.+d}

\maketitle
%

Atomic spin comagnetometers \cite{aleksandrov1983restriction,Lamoreaux_1986} are used in a number of precision fundamental physics experiments \cite{Safranova_2017}. Recent efforts on miniaturization of atomic sensors \cite{Liew_2004,Knappe_2006} have  led to the development of  chip-scale systems for polarization of nuclear spins that have been used  to search for new short-range spin-dependent forces \cite{Bulatowicz_2013},  for inertial rotation sensing \cite{Donley_2009,Donley_2010,Larsen_2012,Walker_2016,Karlen2018}, for magnetometry \cite{LarsenMag} and microfluidic NMR detection \cite{Pines_2014,Kennedy_2017}.  These applications use mm-sized cells containing alkali metals and noble gas isotopes with nuclear spins, such as $^{129}$Xe  or $^{131}$Xe, which are polarized by optical pumping and spin exchange. The sensitivity in such experiments is determined by the nuclear-spin coherence time, which is often dominated by spin interactions with cell walls and is typically short in small cells that have a large surface-to-volume ratio.  Here we demonstrate batch fabrication of stemless anodically bonded cells containing  $^{3}$He  and $^{129}$Xe with nuclear-spin coherence times of 4 hours and 300 s, respectively. This represents a factor of 10 to 100 improvement in the $^{129}$Xe coherence time for micro-fabricated cells and the first detection of $^3$He signals in such cells.
Longer coherence times have been observed in larger cells fabricated with traditional glass-blowing techniques \cite{Anger_2008,Gemmel_2010,Gemmel_2010_2}. 

Batch cell fabrication using anodic bonding allows for excellent control of the cell geometry. Typical glass blown or optically contacted cells have a glass stem for cell filling and sealing \cite{Eklund_2008,Larsen_2012}. Others have made stemless cells by allowing  $^3$He to diffuse through quartz walls at a high temperature \cite{Heil_2016}. The cell shape affects the dipolar magnetic interactions between spin-polarized nuclei \cite{Nacher_1995}. Such interactions cause significant frequency shifts in comagnetometer precision measurements  \cite{Allmendinger_2014} and are subject of some controversy \cite{Romalis_2014}. Here we form stemless cells using a silicon wafer with an array of 2~mm diameter  holes covered on both sides with anodically-bonded aluminosilcate glass plates. By making a slight wedge in the Si wafer we  fabricate  a number of  cylindrical cells in one batch with varying height to diameter aspect ratios. We systematically study nuclear-spin dipolar fields in a new regime where atomic diffusion across the cell is much faster than  the time scales of  long-range dipolar interactions and spin relaxation, unlike previous NMR experiments investigating distant dipolar fields in liquids and gases \cite{Warren_1993,Acosta_2008}. We find that for a certain cylindrical cell aspect ratio the average dipolar fields are eliminated, in good agreement with  theory. An optimal and well-defined geometry will improve stability of nuclear-spin comagnetometers  used for fundamental physics  measurements \cite{Bulatowicz_2013,Glenday_2008,Tullney_2013,Allmendinger_2014,Rosenberry_2001}.  

Control of long-range dipolar fields allows us to resolve a small scalar frequency shift between $^3$He and $^{129}$Xe nuclear spins mediated by a second-order electron Fermi-contact interaction. Such through-space $J$-coupling  in van der Waals molecules has been theoretically studied in NMR \cite{Harris_1998,Pecul_2000,Bagno_2003} and was  first observed  experimentally between  $^{129}$Xe and $^1$H in a liquid mixture of Xe and pentane \cite{Ledbetter_2012}. Here  we report the first observation of spin-spin $J$-coupling between nuclear spins in the gas phase. 

\begin{figure}
\includegraphics[width=3in]{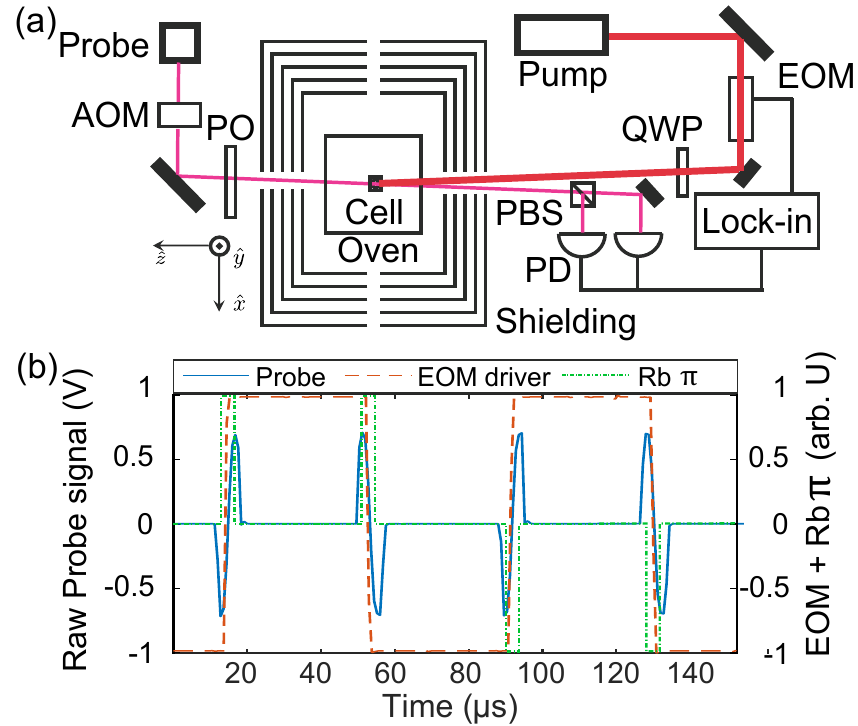}
\caption{(a) Parallel pump-probe pulsed $^{87}$Rb magnetometer with one optical axis along $\hat{z}$. (b) An EOM square wave alternates $\sigma^+$/$\sigma^-$ pump light and $\pi_{\pm y}$ pulses are applied to retain $^{87}$Rb polarization. The probe laser is gated with an AOM to detect only $^{87}$Rb polarization transitions. A small $B_y$ field changes the $^{87}$Rb transition phase and is detected with a lock-in referenced to half the EOM square wave frequency.}
\label{fig:1}
\end{figure}

{\em Detection of nuclear spin signals in mm-sized cells.--}
We use $^{87}$Rb to polarize  nuclear spins by spin-exchange and detect their  precession with an in-situ $^{87}$Rb magnetometer \cite{Sheng_2014, Limes_2017}, which gives a high signal-to-noise ratio because Rb experiences enhanced nuclear spin magnetic fields  during spin-exchange collisions \cite{Schaefer_1989,Romalis_1998,Ma_2011}. The pulse-train  magnetometer described in \cite{Limes_2017} is adapted here for use with cells with a single optical axis by using counter-propagating pump and probe beams (Fig.~\ref{fig:1}a). After an initial pump time of 30-50 s to polarize the nuclear spins along a $\hat{z}$   bias magnetic field $B_0 \approx 5$ mG ,  the polarization of the on-resonant 795 nm pump laser is switched between $\sigma^+$ and $\sigma^-$ light with an  electro-optic modulator (EOM) at 13 kHz. Simultaneously a train of 3 $\mu$s long magnetic field $\pi$ pulses are applied along $\hat{y}$ to flip the $^{87}$Rb polarization back and forth along $\hat{z}$.  The high $\pi$ pulse repetition rate suppresses  spin-exchange relaxation \cite{Limes_2017}. $^{87}$Rb polarization projection on $\hat{z}$ is detected with paramagnetic Faraday rotation of an off-resonant probe beam passing through the cell to a balanced polarimeter. An acousto-optic modulator turns on the probe laser only during the $\pi$ pulses to avoid unnecessary probe broadening during pumping intervals. A weak $B_y$ field  causes an advancement or retardation of the $^{87}$Rb polarization phase during the $\pi$ pulse flip (Fig.~\ref{fig:1}b).   The  polarimeter signal is sent to a lock-in amplifier referenced to half the EOM frequency  such that the lock-in output is proportional to $B_y$.  We obtain a sensitivity of 300 fT/$\sqrt{\text{Hz}}$ in our miniature cells, and are able to operate this scheme with $B_0$ parallel or perpendicular to the cell's optical axis.

We make miniature vapor cells using an anodically bonded glass-Si-glass construction \cite{Liew_2004} in a custom-built system able to fabricate cells containing isotopically enriched alkali metals and noble gases. 
 The Si wafer is 2~mm nominal thickness with a $7\times7$ array of machined  holes with a diameter $d = 2.005\pm0.005$ mm.  We polished one side of a wafer at a small angle to obtain cell height variations of 1.666 mm to 1.988~mm. 
 The wafer is baked under high vacuum inside the fabrication system to remove contaminants, which is crucial to obtain long $^{129}$Xe lifetimes. We close the cells with 0.2 mm thick aluminosilicate glass SD-2 that has low $^3$He permeability \cite{Kitchingglass}.  After anodically bonding glass on one side of the wafer, we distill $99.9\%$ isotopically pure $^{87}$Rb metal and bond the second glass in an atmosphere of 80 torr N$_2$, 6.5 torr $^{129}$Xe, and 1400 torr of $^{3}$He. Heating during anodic bonding can cause buffer gas pressure loss in cells of up to 30$\%$. A cryogenic storage system recaptures the remnant buffer gas mixture for future use.
 
\begin{figure}
\includegraphics[width=3in]{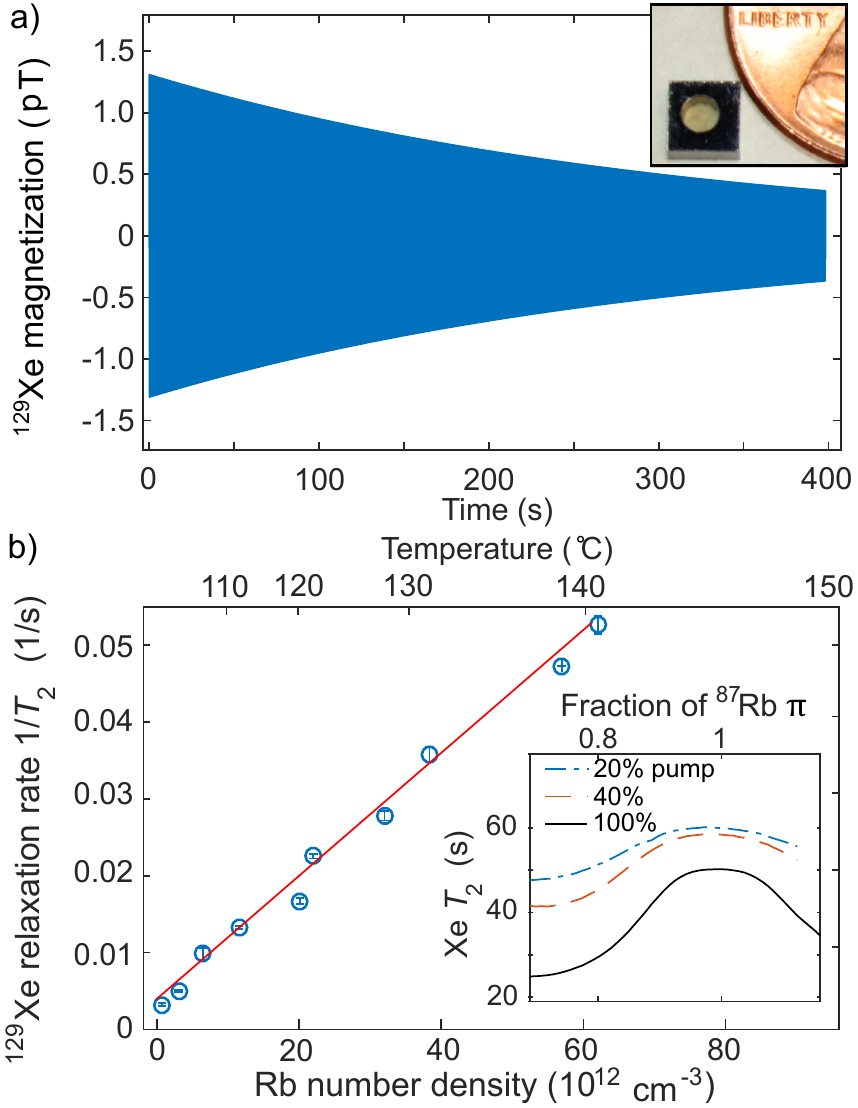}
\caption{a) Spin precession signal from $^{129}$Xe at 73.3$^{\circ}$C with $T_2 = 308$ s. The inset shows a picture of the cell. b)  The dependence of $^{129}$Xe $T_2$ on the Rb number density. The inset shows that at $120^{\circ}$C  $^{87}$Rb pump light intensity and deviations of $\pi$ pulse amplitude from optimal conditions shorten $^{129}$Xe $T_2$ due to Xe diffusion in a $^{87}$Rb polarization gradient.}
\label{fig:2}
\end{figure}

{\em Nuclear spin  relaxation  measurements--}
The dominant sources of $^{129}$Xe spin relaxation are collisions with cell walls and Rb-Xe collisions, while collisional Xe-Xe relaxation \cite{Chann_2002}   is negligible.  After the initial pumping interval a tipping pulse places  $^{129}$Xe spins perpendicular to $B_0$ and their precession is detected with the pulse-train $^{87}$Rb magnetometer. 
A representative signal from the lock-in amplifier is shown in Fig.~\ref{fig:2}a. We fit the data to the function $A\exp(-t/T_2)\sin(\omega_{\text{Xe}} t)$. We show  $^{129}$Xe $1/T_2$  as a function of cell temperature  and Rb density  in Fig.~\ref{fig:2}b and find  $^{129}$Xe wall relaxation time $T_w= 305 \pm 5$~s.
We check the $^{3}$He and $^{87}$Rb densities by measuring the Rb absorption spectrum of the probe laser.  In this cell we find the Rb absorption  FWHM of 23.4~$\pm 1$~GHz corresponding to  $1000$~torr $ ^{3}$He \cite{Romalis_1997}. From the slope in Fig.~\ref{fig:2}b  we find a Rb-Xe spin-exchange rate of $(7.8 \pm 0.7) \!\times\! 10^{-16}$ cm$^3$/s in agreement with a calculation of the spin-exchange rate $7.2 \times 10^{-16}$ cm$^3$/s based on previously measured cross-sections \cite{Nelson_2001, Schrank_2009}. 
%
This indicates $T_w$ does not change significantly over our temperature range. The inset of Fig.~\ref{fig:2}b shows the Rb magnetometer can shorten $ ^{129}$Xe $T_2$ by causing a Rb polarization gradient that $^{129}$Xe diffuses through.  The additional relaxation depends on the $^{87}$Rb pump laser  power and the accuracy of  $^{87}$Rb $\pi$ pulses.  Increasing the $^{87}$Rb $\pi$ pulse repetition rate  decreases this relaxation   because the Rb gradient is reversed more rapidly. For accurate $T_2$ measurements we use proper $\pi$ pulses and low pump power.

{\em $^{3}$He--$^{129}$Xe Interactions.--}
We study magnetic dipolar interactions between $^3$He and $^{129}$Xe spins in the regime where atomic diffusion across the cell is much faster than both the time scale of spin precession in dipolar fields and of spin relaxation. In this regime each spin species has a uniform nuclear magnetization $\bf M$ inside the cell, unlike previous studies of long-range dipolar fields \cite{Grover_1990,Warren_1993}. The spin precession frequencies are determined by the volume average magnetic field inside the cell, which can be calculated using the magnetometric demagnetizing factors $\left \langle H_i\right\rangle_V=-n_i M_i$, $(i=x,y,z$) \cite{Demag}. Analytical expressions for $n_i(\gamma)$ for a cylinder, where $\gamma=h/d$ is the height over diameter, are given in \cite{Joseph_1966,Chen_2006}. The demagnetizing factors satisfy $n_x+n_y+n_z=1$. For a cylinder with $\gamma=0.9065,$ $n_i=1/3$, the same as for a sphere.

The classical average magnetic field needs to be corrected by separating out the contact term $2 \mu_0\delta^3({\bf r}){\bf m}/3$   of the dipolar   field for a point dipole $\bf m$, which can be enhanced or suppressed depending on interactions between atoms that are parametrized by $\kappa$ \cite{Heckman_2003}. We write
\begin{equation}
 \left\langle B^{d}_i\right\rangle_V=\mu_0 \left[ M_i-n_i M_i+\frac{2}{3}(\kappa-1)  M_i \right ].
\end{equation}
Bloch equations describe spin precession in the presence of  a constant bias field and small nuclear-spin dipolar fields. In our case only $^3$He has a significant magnetization. The dipole field experienced by $^{129}$Xe
can be written as $\left\langle B^{\text{Xe}}_z\right\rangle_V=\mu_0 (1/3-n_z +2\kappa_\text{HeXe}/3)M_z^{\text{He}}$, where the $z$ axis is defined by the magnetic field direction. The rotating components of the $^3$He magnetization do not have a net effect on $^{129}$Xe precession frequency to first order in $M_z^{\text{He}}$. In contrast, the $^3$He precession frequency is affected by the secular co-rotating components of the $^3$He dipolar field but is not affected by the scalar contact interaction with $^3$He . One can write \cite{Jeener_1996,Warren_1995}:
\begin{equation}
\frac{\left\langle {\bf B}^{\text{He}}\right\rangle_V}{\mu_0}= 
\left(\frac{n_z}{2}-\frac{1}{6}\right){\bf 
M}^{\text{He}}+\frac{3}{2}\left(\frac{1}{3}-n_z\right) M_z^{\text{He}}\hat{z}.
\end{equation}
The first term on the right hand side does not generate any frequency shift, since it gives $\left\langle {\bf B}^{\text{He}}\right\rangle_V\parallel{\bf M}^{\text{He}}$.  The effective dipolar field responsible for a $^{3}$He  frequency shift  is  given by the second term. It is   3/2 times larger than for $^{129}$Xe and both are proportional to the $M_z^{\text{He}}$ projection.

\begin{figure}
\includegraphics[width=3in]{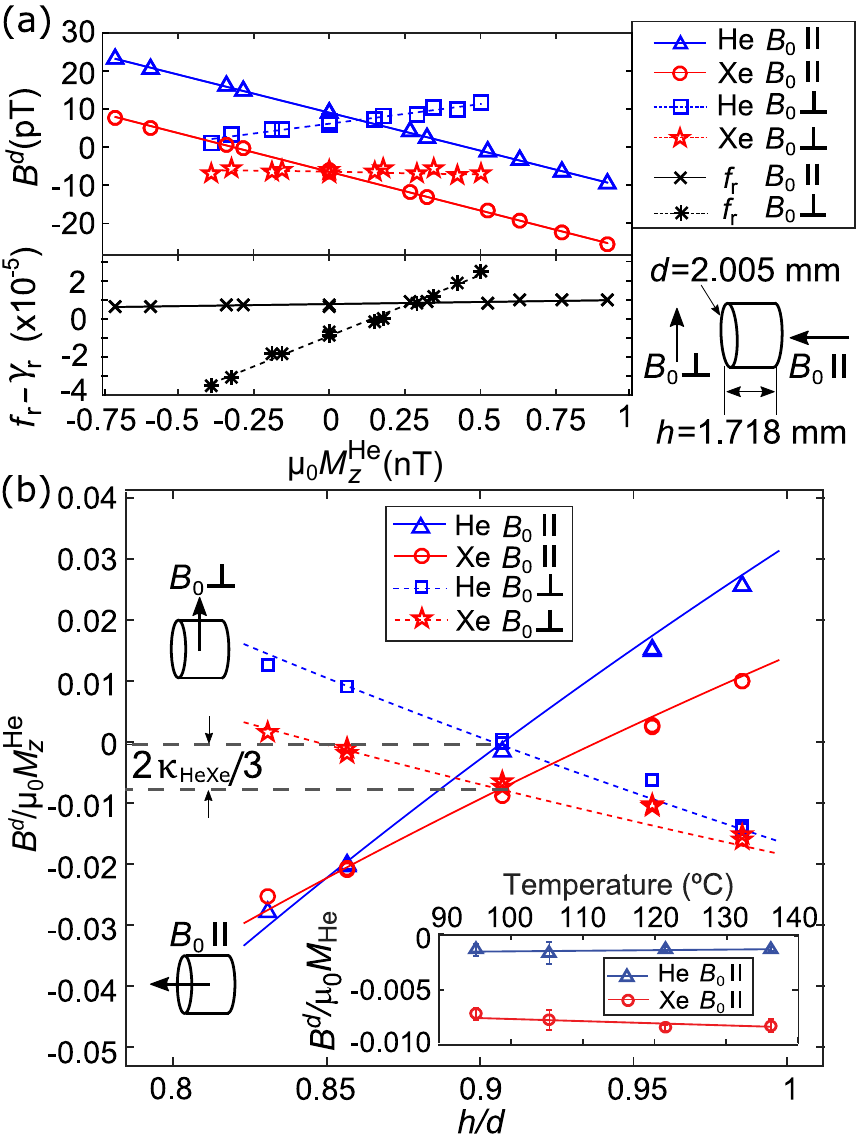}
\caption{(a) The dipolar field $B^{d}$ experienced  by $^{3}$He (triangles) and $^{129}$Xe (circles) from $^{3}$He magnetization $M_{\text{He}}$ along $B_{0}$  $\parallel$ to the cylinder axis and  $^{3}$He (squares) and $^{129}$Xe (stars) for $B_{0}\perp $ to the cylinder axis with $h/d = 0.857$. Lines show linear fits. The comagnetometer frequency ratio $f_r -\gamma_r = \omega_{\text{He}} /\omega_{\text{Xe}}-\gamma_{\text{He}}/\gamma_{\text{Xe}}$ is shown for $B_{0} \parallel$ ($\times$'s) and $B_0 \perp$ ($*$'s). (b) The slope $B^{d}/\mu_{0}M^{\text{He}}_z$ is plotted against the cell aspect ratio $h/d$ for $B_0 \parallel$ and $B_0 \perp$ to the cylinder axis. Lines show theory with one free parameter of  $\kappa_{\text{HeXe}}$.  Inset: Temperature dependence of $\kappa_\text{HeXe}$ for the cell with $h/d = 0.905$.  
}
\label{fig:3}
\end{figure}

We study the effect of dipolar fields in our cells by first polarizing $^3$He for several hours and then creating a small $ ^{129}$Xe polarization. We can also polarize the nuclear spins with $B_0$ perpendicular to the  optical axis by applying fast $\pi$ pulses with $\sigma_+$/$\sigma_-$ pumping that give a time-averaged Rb polarization along $B_0$.
We then apply a tipping pulse calculated to leave a certain percentage of $M_{\text{He}}$  along $B_0$ and place $^{129}$Xe magnetization in the transverse plane. 
Noble-gas precession signals are recorded in a Ramsey-style sequence for about  100 s, with two detection periods separated by  a dark period that has a rotating, two-axis decoupling pulse train applied \cite{Limes_2017}. 
This decoupling pulse train removes noble-gas frequency shifts due to $^{87}$Rb back-polarization and nulls Bloch-Siegert shifts introduced by the pulse train. 
Each detection period is fit to two decaying sine waves to extract the phases with which the noble gases enter and leave the dark period. 
Knowing the number of cycles elapsed during the dark period, we find the in-the-dark free-precession frequencies of the noble gases $\omega_{\text{He}}$ and $\omega_{\text{Xe}}$.
We divide $\omega_{\text{He}}$ and $\omega_{\text{Xe}}$ by their gyromagnetic ratios $\gamma_{\text{He}}$, $\gamma_{\text{Xe}}$ \cite{Makulski_2015} to find the dipolar fields experienced  by $^{3}$He and $^{129}$Xe due to the $\hat{z}$ projection of the $^{3}$He magnetization  $M^{\text{He}}_z$. After the Ramsey measurement we place $M_{\text{He}}$ along or against $B_0$ by using dumping feedback \cite{Alem_2013}.
We avoid systematic errors from $B_0$ drift, chemical shifts, and any effects of remnant $^{129}$Xe polarization projection onto $B_0$ by using identical tipping pulses while alternating the sign of the initial $M_{\text{He}}$ projection along $B_0$. 
$M_{\text{He}}$ is found by comparing the amplitude of the fitted $^{3}$He signal  to the  $^{87}$Rb magnetometer response from a known magnetic field and dividing by $2\mu_0\kappa_0^\text{RbHe}/3 $~
 to account for Rb$-^3$He contact interaction  \cite{Romalis_1998}. We neglect a 1\% correction from the long-range dipolar effect of $M_{\text{He}}$ on $^{87}$Rb. 
The Ramsey sequence is repeated for many $M_{\text{He}}$   values and tipping angles to measure the slope $B^d/\mu_0 M^{\text{He}}_z$. 
\begin{table}

\begin{tabular}{|c|c|c|c|c|}
\hline
~ &$^{129}$Xe field &$^{129}$Xe theory &$^{3}$He field&$^{3}$He theory\\
\hline
$B_0 \parallel$ & $1 $ &$1$  &$1.50\pm0.02$  &$3/2$\\
$B_0 \perp$ & $-0.44 \pm 0.03$  &$-1/2$&  $-0.72 \pm 0.03$&  $-3/4 $\\
\hline
\end{tabular}
\caption{Relative size of the slopes $(B^d/\mu_0M^{\text{He}}_z)/(h/d)$ from fits to Fig.~\ref{fig:3}b scaled to the $^{129}$Xe $B_0 \parallel$ case. }
\label{t:table}
\end{table}

We plot the effective dipolar field experienced by $^{3}$He and $^{129}$Xe due to  $M_z^\text{He}$  for a cell with $h=1.718$ mm, $d=2.005$ mm at 120$^{\circ}$C  in Fig.~\ref{fig:3}a. We repeat the measurements with $B_0$  parallel and perpendicular to the optical axis of the cylindrical stemless cell.  In Fig.~\ref{fig:3}b we plot the slope of the dipolar field  $B^d/\mu_0 M^{\text{He}}_{z}$ as a function of the cell aspect ratio $h/d$ for several cells. Solid lines in Fig.~\ref{fig:3}b represent first-principles calculations of the dipolar fields using the expression for $n_z(\gamma)$. The only free parameter is the value $\kappa_\text{HeXe} = -0.011 \pm 0.001$, where the error is determined by $M_z^\text{He}$  calibration 
uncertainty. The relative size of the long-range dipolar frequency shifts for the four cases are related to each other by simple ratios shown in Table~\ref{t:table}, which are applicable for any cell with uniaxial symmetry \cite{Demag}. 

The existence of a finite $\kappa_\text{HeXe}$ implies that the cell aspect ratio required to cancel   the $^{3}$He  magnetization effect on the frequency ratio in the comagnetometer  is different  from  the condition $n_i=1/3$. We find that in our cell with $h/d = 0.857$ the comagnetometer frequency ratio $f_r$ is insensitive to $M^{\text{He}}_z$ for $B_0 \parallel$ to the cell axis, as shown in the bottom panel of Fig.~\ref{fig:3}a. The temperature dependence of $\kappa_\text{HeXe}$, shown in the inset of Fig.~\ref{fig:3}b is relatively weak. Operation of a $^{3}$He-$^{129}$Xe comagnetometer in a well-defined cylindrical cell with this aspect ratio will improve its long term stability.

The contact spin-spin $J$ couplings for noble gases have been calculated for  $^{129}$Xe-$^{131}$Xe \cite{Vaara_2013} and for $^3$He~\cite{Pecul_2000}. For spin couplings in gases and liquids with fast molecular motion it is more convenient to parametrize  the interaction in terms of $\kappa$ \cite{Heckman_2003,Ledbetter_2012}:
\begin{equation}
\kappa=-\frac{3 \pi}{\mu_0 \gamma_1 \gamma_2 \hbar} \int 4 \pi r^2 g(r)J(r)dr
\end{equation}
 where $g(r)$ is the radial intermolecular distribution function. For example, for the $^{129}$Xe-$^{131}$Xe  calculation  the low gas density limit can be obtained by considering two Xe atoms confined to a spherical  cavity of a certain size \cite{Vaara_2013}. It gives $\kappa_{^{129}\text{Xe}-^{131}\text{Xe}}= - 0.27$, which may 
give observable effects in  $^{129}$Xe-$^{131}$Xe comagnetometers  
\cite{Bulatowicz_2013}. The $^{129}$Xe-$^1$H contact interaction was 
measured and calculated in \cite{Ledbetter_2012}, 
$\kappa_{^{129}\text{Xe}-^{1}\text{H}}= - 0.0014$. The value of $\kappa$ 
increases for heavier atomic pairs, as can be expected.

\begin{figure}
\includegraphics[width=3in]{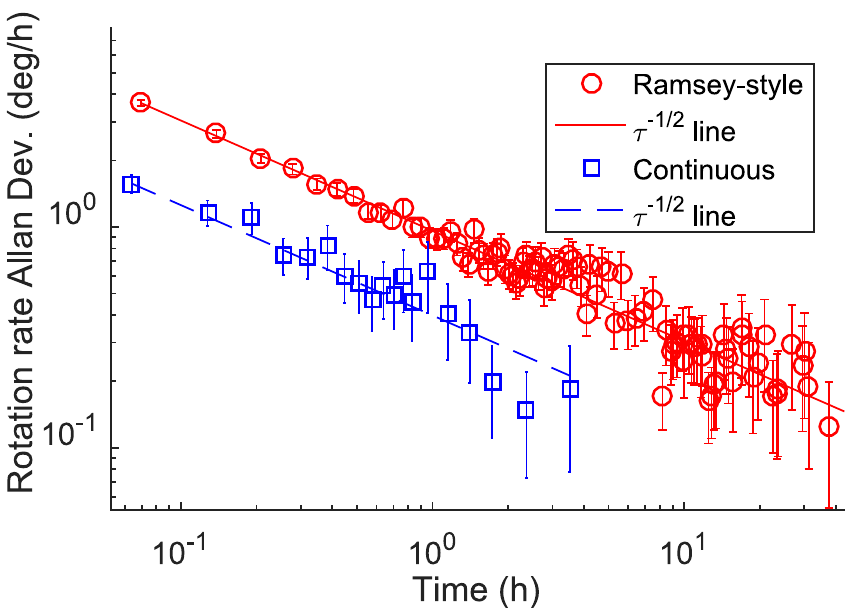}
\caption{Allan deviation for continuous comagnetometer measurements   and Ramsey-style sequence with active Rb depolarization  in a batch-fabricated anodically bonded cell at 120$^{\circ}$C. }
\label{fig:4}
\end{figure}

The cells fabricated for this experiment were used to test the comagnetometer and gyroscope performance of the system, similar to measurements in \cite{Limes_2017}. For a cell  operating at 120$^\circ$C the Cram\'er-Rao frequency uncertainty lower bound using the best signal-to-noise ratio measured for $^{3}$He and $^{129}$Xe signals was found to be  0.005 deg/$\sqrt{\text{h}}$.
We also operated the comagnetometer for an extended period of time to determine its  long-term stability. In  Fig.~\ref{fig:4} we compare the Allan deviation for the  Ramsey-style sequence with active Rb depolarization  and for continuous measurements with the Rb $\pi$-pulse train magnetometer, as described in more detail in \cite{Limes_2017}.   The frequency uncertainty is limited by the $^{3}$He frequency error due to low steady-state $M_{\text{He}}$ at a temperature that allows for long $^{129}$Xe $T_2$. The uncertainty  is about an order of magnitude  worse than for glass-blown cells with 100 times larger volume used in \cite{Limes_2017}. Using microfabricated cells with higher $^3$He pressure would improve the performance.

In conclusion, we demonstrated fabrication of miniature anodically bonded vapor cells containing $^{3}$He, $^{129}$Xe, $^{87}$Rb, and N$_2$ with   long $^{129}$Xe  and $ ^{3}$He coherence times. Precise control over the cell geometry allowed us to make detailed measurements of the  long-range nuclear dipolar fields due to $ ^{3}$He magnetization in the regime of fast atomic diffusion, which are in good agreement with calculations. We observe a small but finite scalar interaction between $ ^{3}$He and $^{129}$Xe. As a result, the optimal cell shape for operation of a dual noble-gas comagnetometer is slightly different from a sphere. A cylindrical geometery can be chosen for a given orientation of the cell with respect to the bias field to cancel dipolar and scalar shifts.

This work was funded by DARPA and NSF.
We would like to thank Juha Vaara for useful discussions.
%

\end{document}